\documentclass[12pt]{article}
\usepackage{graphicx}
\begin{document}
\begin{center}

{\Large \bf Influence of the re-scattering process on polarization observables in reaction 
$ \gamma d \rightarrow p p \pi^{-} $ 
in $\Delta$--resonance  region.\footnote{
This work was supported by Russian Foundation for Basic 
Research   
N 98-02-17993, 
N 98-02-17949 and by grant 
N 96-0424 
from INTAS.}.} \\ 	[5mm]
\end{center}
\begin{center}
\large{
{A.Yu. Loginov, A.A. Sidorov, V.N. Stibunov} \\
{\small \it Nuclear Physics Institute
at Tomsk Polytechnical University, Tomsk, Russia}}\\
[6mm]
\end{center}
\begin{center}
\large
{This work to be published in Yadernaya Fizika.\\

Stend report at the Fourth International Conference\\
 on Physics at storage rings,
 September 121-16, 1999,\\ Bloomington, Indiana, USA.}
\\[5mm]
\end{center} 
\begin{abstract}
The influence of the  pion-nucleon and 
nucleon-nucleon re-scattering effects on the polarization observables of the
reaction $ \gamma d \rightarrow p p \pi^{-} $ in $\Delta $ -- isobar
region is investigated.
Pion-nucleon and nucleon-nucleon re-scattering are studied in the diagrammatic
approach.
Relativistic-invariant forms of the photoproduction and pion-nucleon 
scattering operators are used.
The unitarization procedure in $K$-matrix approach is applied for the 
resonance partial amplitudes.
It is shown a considerable influence of the re-scattering on the polarization
observables of this reaction  in 
$\Delta$ -- resonance region for large momenta of the final protons.
\end{abstract}

\section{}  
\hspace*{\parindent} 
\hspace*{\parindent} 	   
Recently the experiment on studying the reaction 
$e\vec{d} \rightarrow e'pp\pi^{-}$ at photon point on the internal 
polarized deuterium target at the VEPP-3 storage ring has been performed
\cite{sid, osip}.
In the experiment two protons with momenta exceeding $ 300 \ MeV/c$
were detected in coincidence.
The obtained differential cross sections and asymmetry components,
especially ${\displaystyle a_{20}}$, differ considerably from those
calculated in the spectator model.
In such a situation it seems to be reasonable to take into account the Final 
State Interaction (FSI) of the reaction, namely pion-nucleon and 
nucleon-nucleon re-scattering.
In the work \cite{lag1} the spin-averaged differential cross-section of the
$\pi^{-}$-photoproduction reaction on the deuteron 
\begin{equation}
\label{eq1}
\gamma d \rightarrow p p \pi^{-} \ .
\end{equation} 
was calculated in diagrammatic approach with an account of FSI.
The contribution of FSI to the polarization observables in deuteron 
electro- and photo-disintegration reactions are calculated in Refs. 
\cite{rec1,lev}.
In present work an influence of FSI on the polarization observables of the 
reaction (1) is studied. 
The results of our calculations can be used as a basis for the simulation
of the behavior of polarization observables 
for specified phase-space elements of the reaction (1), 
for comparison with existent experimental data and for an optimal planning 
of new polarization experiments with large momenta of detected protons.
Such a comparison can help to determine whether it is necessary to include 
into consideration more complicated mechanisms of the reaction (1).

A description of the used model and the expressions for the amplitudes of
various re-scattering mechanisms are given in section 2.
Section 3 is devoted to  the determination of the helicity amplitudes and
polarization observables.
The results of the calculations are presented in section 4.

\section{}
\hspace*{\parindent} 
To derive an amplitude for the reaction (1) we will use a diagrammatic
approach developed in the works \cite{shap1,shap2}.
The diagrammatic approach in the analysis of the $\pi$--meson
photo-production reaction on the deuteron was applied in Refs. 
\cite{lag1, blom}.
We will take into consideration the contribution of the diagrams shown in
Fig.1.

The first diagram corresponds to the spectator model.
This model sufficiently well describes experimental data for kinematics
close to the quasi-free $\pi$--meson photoproduction.
In the case of sufficiently large momenta of both nucleons (above 200 $MeV/c$) 
one must take into account contributions of next diagrams which describe 
final state re-scattering.

\begin{figure}[htb]
\centering \includegraphics[width=0.8\textwidth]{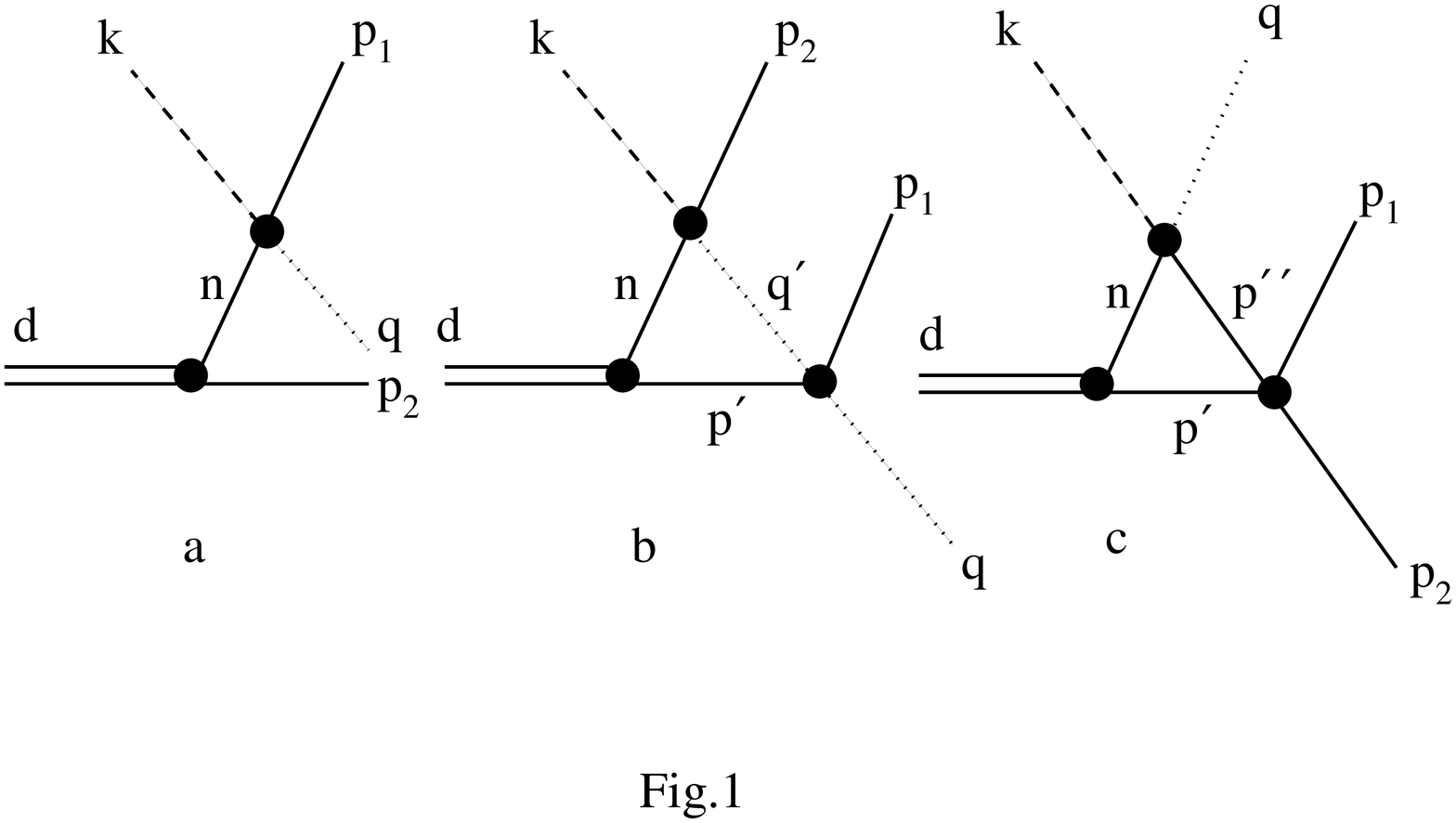}
%\includegraphics[angle=-90]
%\caption{} \label{fig1.}
\end{figure}

\suppressfloats[no-loc]

Qualitatively explaining, this is because a probability to find a 
nucleon-spectator in the deuteron reduces with an increase of its momentum.
In the spectator model this leads to a reduction of the reaction amplitude.
In the same time the re-scattering amplitude is determined by an integral
over nucleon momentum in the deuteron.
The main contribution to this integral even for large nucleon momenta in the
final state may come from a low-momentum part of the deuteron wave function.
That is why the role of re-scattering effects increases with nucleons momenta.
  
A non-relativistic form of the photoproduction amplitude relevant in the 
$\Delta$--isobar region in an arbitrary frame for small $p^{2}/m^{2}$, where
$p$ and $m$ being nucleon momentum and nucleon mass, was used 
in \cite{lag1, blom} to construct analytical expressions corresponding 
to the diagrams depicted in Fig.1.
For a description of the resonance partial amplitude \ $M^{\frac{3}{2}}_{1+}$
the $\Delta$--mass was taken with an imaginary component, which is a function
of isobar energy.
The used parametrization of this function did not allow to describe the
photoproduction of both charged and neutral $\pi$--mesons in a unified way.
We used the amplitude of $\pi$--meson on nucleon photoproduction from 
\cite{david}. 
This amplitude is written in relativistic-invariant and gauge-invariant form, 
it realizes a pseudovector variant of $\pi N$--interaction and takes into
account a contribution of Born diagrams in $s-$, $t-$ and $u-$ channels,
as well as contributions of $s-$ and $u-$ channels of $\Delta$--isobar and
of $t-$ channel exchange of $\omega-$ and $\rho-$ mesons.
For the resonance partial amplitudes $E^{\frac{3}{2}}_{1+}$  and
$ M^{\frac{3}{2}}_{1+}$ a procedure of unitarization in $K$--matrix approach
was applied.
This allowed to avoid an introduction of the additional imaginary component
to the $\Delta$--isobar mass and permitted to describe the process of 
photoproduction of charged and neutral $\pi$--mesons in a unified way.

Pion-nucleon scattering was described by a relativistic-invariant amplitude,
given in \cite{olsson}, where also a pseudovector variant of $\pi N$ 
interaction was realized.
In that paper the contributions of Born parts in $s-$, $t-$, $u-$ channels,
$\Delta$--isobar in $s-$ and $u-$ channels, $t$--channel exchange of
$\sigma-$ and $\rho-$ mesons were accounted for. 
For the resonance partial-wave amplitude $P_{33}$ a unitarization procedure
in a $K$--matix approach was used in a similar way as in the case of the
photoproduction of $\pi$--mesons on nucleons.

Nucleon-nucleon scattering amplitude was presented in a form of 
multipole expansion with partial waves of upto $L=2$ orbital momentum. 
Partial phase shifts of nucleon scattering were taken from \cite{mac}. 
	
The wave function of the final $NN$--state in a coupled basis in spin and 
isospin spaces which satisfies the symmetry rules with respect to a 
permutation of identical nucleons has the form:
\begin{equation}
\left| p_{1},p_{2},sm_{s},tm_{t} \right\rangle = \frac{1}{\sqrt{2}} \left(\left| p_{1} \right\rangle^{1}
\left| p_{2} \right \rangle^{2} + \left(-\right)^{1+s+t} \left| p_{2} \right\rangle^{1}\left| p_{1} \right \rangle^{2}
 \right) \left|s m_{s} \right\rangle \left| t m_{t} \right\rangle \ , 
 \end{equation}
where $p_1$, \ $p_{2}$ -- nucleons momenta, $ s,\, m_{s}$ -- spin of nucleon 
pair and its projection on $z$-axis, $ t,\, m_{t}$ -- isospin of nucleon pair
and its projection on $z$-axis.
In this case the amplitude of the reaction (1) in the spectator model
in a coordinate system where $z$-axis is along the direction of 
$\gamma$--quantum and $d$ is deuteron momentum has the form \cite{shmidt}:
\begin{eqnarray}
 T^{spect}\left(p_{1}, p_{2}, q, s,m_{s} ; k,\lambda_{\gamma}, d,m_{d} \right)  = \phantom{iiiiiiiiiiiiiiiiii} & & \nonumber \\
 = -\sum \limits_{m^{\prime}_{s}} \left\langle s m_{s} \right| 
\left[ \sqrt{\frac{E_{p_{2}}}{E_{d-p_{2}}}} T^{1}_{\gamma n \rightarrow p \pi^{-}}
\left(p_{1},q;d-p_{2},k,\lambda_{\gamma}\right)\Psi_{m^{\prime}_{s},m_{d}}\left(\frac{1}{2}\left(
d-2p_{2} \right) \right) + \right.  & & \nonumber \\
 + \left. (-)^{s} \sqrt{\frac{E_{p_{1}}}{E_{d-p_{1}}}} T^{1}_{\gamma n \rightarrow p \pi^{-}}
\left(p_{2},q;d-p_{1},k,\lambda_{\gamma}\right) \Psi_{m^{\prime}_{s},m_{d}}\left(\frac{1}{2}\left(
d-2p_{1} \right) \right) \right] \left|1 m^{\prime}_{s}\right\rangle \ .  & &
\end{eqnarray}
Here $ p_{1}, \  p_{2} $ and $q$ are the momenta of final nucleons and 
$\pi$--meson, 
$k$, $\lambda_{\gamma}$ -- momentum and helicity of $\gamma$--quantum,
$d$, $m_{d}$ -- deuteron momentum and $z$--projection of deuteron spin, 
$s$, $m_{s}$ -- spin of the nucleon pair in final state and its projection
on $z$-axis,
$m'_{s}$ --  projection on $z$-axis of the total spin of nucleons in deuteron,
$E_{p}$ -- ``on-shell'' energy of a nucleon having a momentum $p$. \
$T^{1}_{\gamma n \rightarrow p \pi^{-}}\left(p_{1},q ;
d-p_{2},k,\lambda_{\gamma}\right)$ -- $\pi^{-}$--meson on neutron
photoproduction amplitude, which is considered as an operator acting on 
spin variables of a first nucleon in a two-nucleon system.
The terms $ \Psi_{m^{\prime}_{s},m_{d}}\left( p \right) $ are 
deuteron formfactors in a coupled-basis presentation. They can be expressed
through the $S$-- and $D$-- deuteron wave-functions as follows \cite{shmidt}:
\begin{equation}
\Psi_{m^{\prime}_{s},m_{d}}\left( p \right) = \left( 2\pi \right)^{\frac{3}{2}}\sqrt{2E_{D}}\sum_{L=0,2}\sum_{m_{L}}
i^{L}u_{L}\left( p \right)Y_{Lm_{L}}\left(\hat{p}\right)\langle L m_{L} 1 m^{\prime}_{s} | 1 m_{d} \rangle \ ,
\end{equation}
where $E_{D}$ -- deuteron energy, $\hat{p}$ -- a unitary vector in the
direction of the momentum $p$.
In present work the deuteron wave functions of Bonn potential (full model)
\cite{bonn} were used in calculations.
Pion-nucleon re-scattering is described by the diagram in Fig.~1b
and the one with identical nucleons in final state been permutated.
For the reaction (1) there is a channel of pion-nucleon 
re-scattering without charge-exchange:
$ \pi^{-}p \rightarrow \pi^{-}p $,  
and the one with charge-exchange:
$\pi^{0}n \rightarrow \pi^{-}p$. 
Taking this into account and using the identity of final nucleons one can
express the contribution of pion-nucleon re-scattering into the amplitude
of the reaction (1) as follows:
\begin{eqnarray}
 T^{\pi N}\left(p_{1}, p_{2}, q, s,m_{s} ; k,\lambda_{\gamma}, d,m_{d} \right) = 
\phantom{iiiiiiiiiiiiiiiiiiiiiiiiiiiiiiiii} & & \nonumber \\ 
\int \frac{d^{4}p^{\prime}}{\left( 2 \pi \right)^{4}}\sum_{m^{\prime}_s} 
\frac{\left \langle s m_{s} \right| \left[ T^{1}_{\pi N}
\left( p_{1},q;p^{\prime},q^{\prime} \right) T^{2}_{\gamma N}\left( p_{2},q';d-p^{\prime},k
,\lambda_{\gamma}\right) \right] \left|1 m^{\prime}_{s}\right\rangle}
{\left( n^{0}-E_{d-p^{\prime}}+i\epsilon \right)\left( p^{\prime 0}-E_{p^{\prime}}+i\epsilon \right)
\left(q^{\prime 2}-m_{\pi}^{2}+i\epsilon\right)} 
V\left(d-p^{\prime},p^{\prime};m^{\prime}_{s},m_{d}\right)+ & & \\
+(-)^{s}\int \frac{d^{4}p^{\prime}}{\left( 2 \pi \right)^{4}}\sum_{m^{\prime}_s} 
\frac{\left \langle s m_{s} \right| \left[ T^{1}_{\pi N}
\left( p_{2},q;p^{\prime},q^{\prime} \right) T^{2}_{\gamma N}\left( p_{1},q';d-p^{\prime},k
,\lambda_{\gamma}\right) \right] \left|1 m^{\prime}_{s}\right\rangle}
{\left( n^{0}-E_{d-p^{\prime}}+i\epsilon \right)\left( p^{\prime 0}-E_{p^{\prime}}+i\epsilon \right)
\left(q^{\prime 2}-m_{\pi}^{2}+i\epsilon\right)} 
V\left(d-p^{\prime},p^{\prime};m^{\prime}_{s},m_{d}\right) \ , & & \nonumber
\end{eqnarray}  
where
$ T^{1}_{\pi N}\left( p_{1},q;p^{\prime},q^{\prime} \right) $  -- 
pion-nucleon scattering amplitude acting as an operator on spin variables 
 $|1 m'_{s} \rangle$ and $|s m_{s} \rangle$ of a first nucleon in a 
two-nucleon system,
 \\ $T^{2}_{\gamma N}\left(p_{2},q';d-p^{\prime},k,\lambda_{\gamma}\right)$ -- 
$\pi$--meson photoproduction amplitude acting as an operator on spin 
variables $|1 m'_{s} \rangle$ ¨ $|s m_{s} \rangle$ of a second nucleon 
in a two-nucleon system.
$n^{0},\ p'^{0}$ ¨ $ E_{d-p'}, \ E_{p'}$ are ``off-shell'' and ``on-shell''
energy of nucleons with momenta $d-p'$ ¨ $p'$ respectively.
The term $[T^{1}_{\pi N}T^{2}_{\gamma N}]$ is related to the sum
over isospin variables in the two-particle operator:
 \begin{eqnarray}
 [T^{1}_{\pi N}T^{2}_{\gamma N}]=T^{1}_{\pi^{-} p \rightarrow \pi^{-} p}T^{2}_{\gamma n \rightarrow p \pi^{-}} - 
 T^{1}_{\pi^{0}n \rightarrow \pi^{-}p}T^{2}_{\gamma p \rightarrow p \pi^{0}} \ .
 \end{eqnarray}
Due to isospin-0 of the deuteron the amplitudes of pion-nucleon scattering
with and without the charge-exchange contribute to the re-scattering amplitude
with opposite signs.
The term $V\left(p^{\prime \prime},p^{\prime};m^{\prime}_{s},m_{d}\right)$ 
is a $ Dnp $--vertex function which in non-relativistic limit is connected to
$ \Psi_{m^{\prime}_{s},m_{d}}$ by \cite{lag1}:
 \begin{eqnarray}
 V\left(p^{\prime \prime},p^{\prime};m^{\prime}_{s},m_{d}\right) = \left( E_{D}-E_{p^{\prime}}
 -E_{p^{\prime \prime}} \right)\Psi_{m^{\prime}_{s},m_{d}} 
 \left(\frac{1}{2}\left(p^{\prime \prime}-p^{\prime}\right)\right) \ ,
 \end{eqnarray}
where $ p', \  p''$ are momenta of nucleons in deuteron.
In the expression (5) relativistic nucleon propagators were first 
expressed as a sum of the terms corresponding to virtual nucleons 
with positive and negative energy and then only the ones with positive 
energies were remained.
This is because the influence of virtual nucleons with negative energy 
starts to manifest itself at momenta $\sim 1 $ GeV \cite{horn}, while such
proton momenta in the kinematic region of $\Delta$--isobar are not reached.
One has to use relativistic form for the pion propagator in the kinematic
region of $\Delta$--isobar.
The integrand has four poles of a variable $p^{\prime 0}$.
Two poles are in the upper half-plane of complex variable $ p^{\prime 0} $,
the other two are in the lower half-plane:
 \begin{eqnarray}
 & p^{\prime 0}_{1+}=E_{D}-E_{d-p'}+i\epsilon , \phantom{iiiiiiiiiiiiiiiiiiii}
   p^{\prime 0}_{2+}=p^{0}_{\Delta}-\omega_{q \prime}+i\epsilon, &   \\
 & p^{\prime 0}_{1-}=E_{p \prime}-i\epsilon, \phantom{iiiiiiiiiiiiiiiiiiiiiiiiiiiiii}
 p^{\prime 0}_{2-}=p^{0}_{\Delta}+\omega_{q \prime}-i\epsilon \ , & \nonumber
 \end{eqnarray}
where $p^{0}_{\Delta}$ is the energy of a $\pi N$--pair participating in the
re-scattering, $\omega_{q'}$ -- the  ``on shell'' energy of a $\pi$--meson
with a momentum $q'$. 
Integration over energy in (5) is done by closing up a contour in the
lower half-plane, only a residue in a nucleon pole $p^{\prime 0}_{1-}$
is considered, while a residue in the pion propagator $p^{\prime 0}_{2-}$
can be neglected due to its smallness \cite{lag1}.
When integrating over 3-momentum of the nucleon the pion propagator is
taking as:
 \begin{equation}
 \frac{1}{q^{\prime 2}-m^{2}_{\pi}+i\epsilon}=P\frac{1}{q^{\prime 2}-m^{2}_{\pi}}-i\pi\delta\left(
 q^{\prime 2}-m^{2}_{\pi} \right)\ , 
 \end{equation}
and the expression (5) is expanded into a sum of terms corresponding
to the contribution of $\delta$ -- function and the main value of the 
integral: 
 \begin{eqnarray}
 T^{\pi N}\left(p_{1}, p_{2}, q, s,m_{s} ; k,\lambda_{\gamma}, d,m_{d} \right) = \phantom{jjjjjjjjjjjjjjjjjj} & & \\
 T^{\pi N}_{on}\left(p_{1}, p_{2}, q, s,m_{s} ; k,\lambda_{\gamma}, d,m_{d} \right) +
 T^{\pi N}_{off}\left(p_{1}, p_{2}, q, s,m_{s} ; k,\lambda_{\gamma}, d,m_{d} \right) , \nonumber & & 
 \end{eqnarray}
where :
 \begin{eqnarray}
 T^{\pi N}_{on}\left(p_{1}, p_{2}, q, s,m_{s} ; k,\lambda_{\gamma}, d,m_{d} \right)= \phantom{iiiiiiiiiiiiiiiiiiiiiiiiii} & & \\
 = \frac{-1}{16 \pi^{2} \left|p_{\Delta}\right|} \int\limits_{0}^{2 \pi} d \phi^{\prime}
 \int\limits_{\left|p_{-}\right|}^{p_{+}}p^{\prime}d p^{\prime}\sum_{m^{\prime}_s} \left \langle s m_{s} \right|
 \left[ T^{1}_{\pi N}\left( p_{1},q;p^{\prime},q^{\prime} \right)
 T^{2}_{\gamma N}\left( p_{2},q';d-p^{\prime},k,\lambda_{\gamma}\right) \right] \left|1 m^{\prime}_{s}\right\rangle 
 \times & & \nonumber \\
 \times \Psi_{m^{\prime}_{s},m_{d}} \left(\frac{1}{2}\left( d-2p^{\prime} \right)\right)+
\phantom{iiiiiiiiiiiiiiiiiiiiiiiiiii}
 & & \nonumber \\
 +\frac{(-)^{1+s}}{16 \pi^{2} \left|p_{\Delta}\right|} \int\limits_{0}^{2 \pi} d \phi^{\prime}
 \int\limits_{\left|p_{-}\right|}^{p_{+}}p^{\prime}d p^{\prime}\sum_{m^{\prime}_s} \left \langle s m_{s} \right|
 \left[ T^{1}_{\pi N}\left( p_{2},q;p^{\prime},q^{\prime} \right)
 T^{2}_{\gamma N}\left( p_{1},q';d-p^{\prime},k,\lambda_{\gamma}\right)\right]\left|1 m^{\prime}_{s}\right\rangle 
 \times & & \nonumber \\
 \times \Psi_{m^{\prime}_{s},m_{d}} \left(\frac{1}{2}\left( d-2p^{\prime} \right)\right) ,
\phantom{iiiiiiiiiiiiiiiiiiiiiiiiiiiii}
 & & \nonumber
 \end{eqnarray}
  
\begin{eqnarray}
 T^{\pi N}_{off}\left(p_{1}, p_{2}, q, s,m_{s} ; k,\lambda_{\gamma}, d,m_{d} \right)= 
 \phantom{iiiiiiiiiiiiiiiiiiiiiiiiiiiiiiiiiiiiiiii} & &  \\
 -i P \int \frac{d^{3}p'}{2\pi^{3}} \sum_{m^{\prime}_s} \frac { \left \langle s m_{s} \right|
 \left[T^{1}_{\pi N}\left( p_{1},q;p^{\prime},q^{\prime} \right)
 T^{2}_{\gamma N}\left( p_{2},q';d-p^{\prime},k, \lambda_{\gamma}\right)\right]
 \left|1 m^{\prime}_{s}\right\rangle}{q^{\prime 2}-m^{2}_{\pi}} 
 \Psi_{m^{\prime}_{s},m_{d}} \left(\frac{1}{2}\left( d-2p^{\prime} \right)\right)+
% \Bigg{|}_{{\textstyle p^{\prime 0}=E_{p\prime}}}
 & & \nonumber \\
+(-)^{1+s}i P \int \frac{d^{3}p'}{2\pi^{3}} \sum_{m^{\prime}_s} \frac { \left \langle s m_{s} \right|
 \left[ T^{1}_{\pi N}\left( p_{2},q;p^{\prime},q^{\prime} \right)
 T^{2}_{\gamma N}\left( p_{1},q';d-p^{\prime},k, \lambda_{\gamma}\right)\right]
 \left|1 m^{\prime}_{s}\right\rangle}{q^{\prime 2}-m^{2}_{\pi}} 
 \Psi_{m^{\prime}_{s},m_{d}} \left(\frac{1}{2}\left( d-2p^{\prime} \right)\right) \ . 
% \Bigg{|}_{{\textstyle p^{\prime 0}=E_{p\prime}}}
 & & \nonumber 
 \end{eqnarray}
In the expression (11) the terms $p_{-}$  and $ p_{+}$ are given by:
 \begin{eqnarray}
 p_{\pm} = \frac{|P_{\Delta}|}{Q}E_{c.m.} \pm \frac{P^{0}_{\Delta}}{Q}|p_{c.m.}|  \ , & &
 \end{eqnarray}
where $Q$ is an invariant mass of $\pi N $ pair, $E_{c.m.},\ p_{c.m.}$ -- 
nucleon energy and momentum in the $\pi N $ center-of-mass frame,
$P^{0}_{\Delta},\ P_{\Delta}$ -- energy and momentum of $\pi N $ pair in
a used frame.
$|p_{-}|$ and $ p_{+}$ are minimal and maximal nucleon momentum in a frame
where energy and momentum of $\pi N $ -- pair are $P^{0}_{\Delta},P_{\Delta}$.
Note that the values of $p_{+},\  p_{-}$ and kinematic variables in the 
expression (13) are different for the first and second terms of the 
expression (11).

In the amplitude (11) the integration over momentum $p'$ is carried out in a 
frame where $z$--axis directed along the momentum of $\pi N$ -- pair,
and the integration over $ \cos(\theta')$ allows to get rid of 
$\delta$ -- function and to fix the angle $\theta'$.
 
The magnitude of the amplitude (11) strongly depends on an integration
limit $|p_{-}|$.
If a nucleon motion in the deuteron and  the 
contribution of the deuteron $D-$state wave function are not included 
into consideration, then each term 
in the amplitude (11) in the Laboratory frame are proportional to the 
integral:
\begin{eqnarray}
  I_{lab.} = \int\limits_{\left|p_{-}\right|}^{p_{+}}p'dp'u_{0}(p')  \  \  . & & 
\end{eqnarray}   
In the kinematic region of $\Delta$ -- isobar an upper limit of $p_{+}$
nearly everywhere exceeds $300 \ MeV/c$, while a lower limit $|p_{-}|$
strongly depends on the momenta of nucleons and $\pi$--meson and varies in
a range from $0$ to $300 \ MeV/c $.
Therefore the integral (14) strongly increases at small $|p_{-}|$, while
weakly depends on $p_{+}$.
In the center-of-mass frame of the reaction (1) an azimuthal dependence 
appears in an argument of the deuteron wave function, and each term
in the amplitude (11) turns out to be proportional to the integral:

 \begin{eqnarray}
I_{c.m.} = \int\limits_{0}^{2 \pi} d \phi'\int\limits_{\left|p_{-}\right|}^{p_{+}}p'dp'u_{0}(\frac{1}{2}(|d-2p'|)) \ ,
\end{eqnarray}  
which also depends mainly on the lower integration limit $|p_{-}|$.
    
In the amplitudes $T^{\pi N}_{on}$ and $T^{\pi N}_{off}$ the residue in 
nucleon pole fixes one of nucleons on the mass shell.
The second nucleon is not on the mass shell, however in the used kinematic
region the shift from the mass shell is small.
In the amplitude $T^{\pi N}_{on}$ the relation (9) fixes the $\pi$--meson 
on the mass shell.
In the amplitude $T^{\pi N}_{off}$ the $\pi$--meson is not on the mass shell,
but the major contribution to the integral comes from the region where 
the ``on-shell'' amplitude of $\pi$--meson photo-production and that of 
pion-nucleon scattering are applicable.
The amplitudes (11), (12) were obtained by numerical integration. 
In the approximation used for the calculation of the main value of the 
integral the photo-production and pion-nucleon scattering amplitudes 
were factored out of integral for zero-momenta of nucleons inside the 
deuteron.

The diagram in Fig 1c and the same one but with identical nucleons 
permutated in final state correspond to the nucleon-nucleon re-scattering.
The contributions of these two diagrams to the amplitude of the 
$NN$--re-scattering are equal in magnitude and have opposite signs.
Since diagrams with permutated fermions contribute with opposite signs 
to the amplitude, these contributions are added and the expression for
the $NN$--re-scattering amplitude in a coupled-basis approach is written as:

\begin{eqnarray}
 & T^{N N}\left(p_{1}, p_{2}, q, s,m_{s} ; k,\lambda_{\gamma}, d,m_{d} \right) =  & 
\end{eqnarray}
\begin{eqnarray} 
 2\int \frac{d^{4}p^{\prime}}{\left( 2 \pi \right)^{4}}\sum_{ \stackrel {\scriptstyle s'' m''_{s}}{ m^{\prime}_s}} 
\frac{ \left \langle s m_{s} \right| T^{1}_{pp \rightarrow pp}\left( p_{1},p_{2};p^{\prime},p^{\prime \prime} \right)
|s''m''_{s} \rangle \langle s''m''_{s}|
T^{2}_{\gamma n \rightarrow p \pi^{-}}\left( p^{\prime \prime},q;d-p^{\prime},k,
\lambda_{\gamma}\right)
\left|1 m^{\prime}_{s}\right\rangle}
{\left( n^{0}-E_{d-p^{\prime}}+i\epsilon \right)\left( p^{\prime 0}-E_{p^{\prime}}+i\epsilon \right)
\left( p^{\prime \prime 0}-E_{p^{\prime \prime}}+i\epsilon \right) }  \times \nonumber & & \\
 \times V\left(d-p^{\prime},p^{\prime};m^{\prime}_{s},m_{d}\right) \ , \phantom{mmmmmmmmmmmmmmmmmm}
\nonumber  & &  
\end{eqnarray}
where $T^{1}_{pp \rightarrow pp}\left( p_{1},p_{2};p^{\prime},
p^{\prime \prime} \right)$ is the proton-proton scattering amplitude, which
acts as an operator on spin variables of both nucleons in two-nucleon systems
$|s m_{s}\rangle$ and $|s'' m''_{s}\rangle$, \\
$T^{2}_{\gamma n \rightarrow p \pi^{-}}\left( p^{\prime \prime},q;d-p^{\prime},k,\lambda_{\gamma}\right)$ 
is the $\pi$--meson photoproduction amplitude, acting as an operator on 
spin variables of a second nucleon in a two-nucleon system
$|1m'_{s}\rangle$ ¨ $|s''m''_{s}\rangle$.  
The integration over $p^{\prime 0}$ is performed like it was done in case of 
$\pi N$ -- re-scattering by closing a loop in a lower half-plane and evaluating 
a residue in the nucleon pole $p^{\prime 0}=E_{p\prime}-i\epsilon$.
After the photoproduction and $NN$--scattering amplitudes are factored out 
of the integral and a contribution of $D$--state deuteron wave function  
is neglected the nucleon-nucleon re-scattering amplitude can be expressed as
\cite{lag1}: 
\vspace {1cm}
\begin{eqnarray}
T^{N N}\left(p_{1}, p_{2}, q, s,m_{s} ; k,\lambda_{\gamma}, d,m_{d} \right) = 
\phantom{iiiiiiiiiiiiiiiii} & &  \\
 -i \frac{4\pi W}{m p_{c.m.}} \sum_{m_{n} m_{p^{\prime}} m_{p^{\prime \prime}}} 
\langle \frac{1}{2} m_{n} \frac{1}{2} m_{p^{\prime}} | 1 m_{d} \rangle 
\langle \frac{1}{2} m_{p^{\prime \prime}} \frac{1}{2} m_{p^{\prime}} | s m_{s} \rangle 
T_{pp \rightarrow pp}(p_{1},p_{2},s,m_{s} ; \frac{d}{2},p'',s,m_{s}) \times & & \nonumber \\
\times T_{\gamma n \rightarrow p \pi^{-}}( p'',m_{p''},q \, ; \frac{d}{2},m_{n},k, 
\lambda_{\gamma}) 
\int\frac{d^{3}\xi}{\left(2\pi\right)^{3}} 
\frac{(2 \pi)^{\frac{3}{2}}\sqrt{2E_{D}} u_{0}\left(\left|\xi+\frac{P}{2}\right|
\right)}{\sqrt{4\pi}}\frac{\left(p^{2}_{c.m.}+\beta^{2}\right)}{\left(p^{2}_{c.m.}-\xi^{2}+i\epsilon \right) \left(\xi^{2}+
\beta^{2}\right)}  \phantom{l} \ , & & \nonumber 
\end{eqnarray}
where $ m_{n},\ m_{p'},\ m_{p''} $ are  $z$-projections of spins of 
intermediate state nucleons, \ $W$ -- energy of nucleon pair in
center-of-mass frame, $p_{c.m.}$ -- nucleon momentum in the same frame,
$P=p_{1}+p_{2}-d$, \ $p''=p_{1}+p_{2}-\frac{d}{2}$, \ $ \beta=241 \ MeV $. 
The terms $T_{pp \rightarrow pp}(p_{1},p_{2},s,m_{s} ; p',p'',s,m_{s})$
are diagonal matrix elements of the $pp$--scattering amplitude in the 
channel-spin representation, they are expressed as a multipole expansion
through phases of nucleon-nucleon scattering. The integral over $\xi$ 
in (17) is evaluated analytically \cite{lag1}.

\begin{figure} [ht]
\unitlength=1cm
\hbox{
\centering \includegraphics[width=1.0\textwidth]{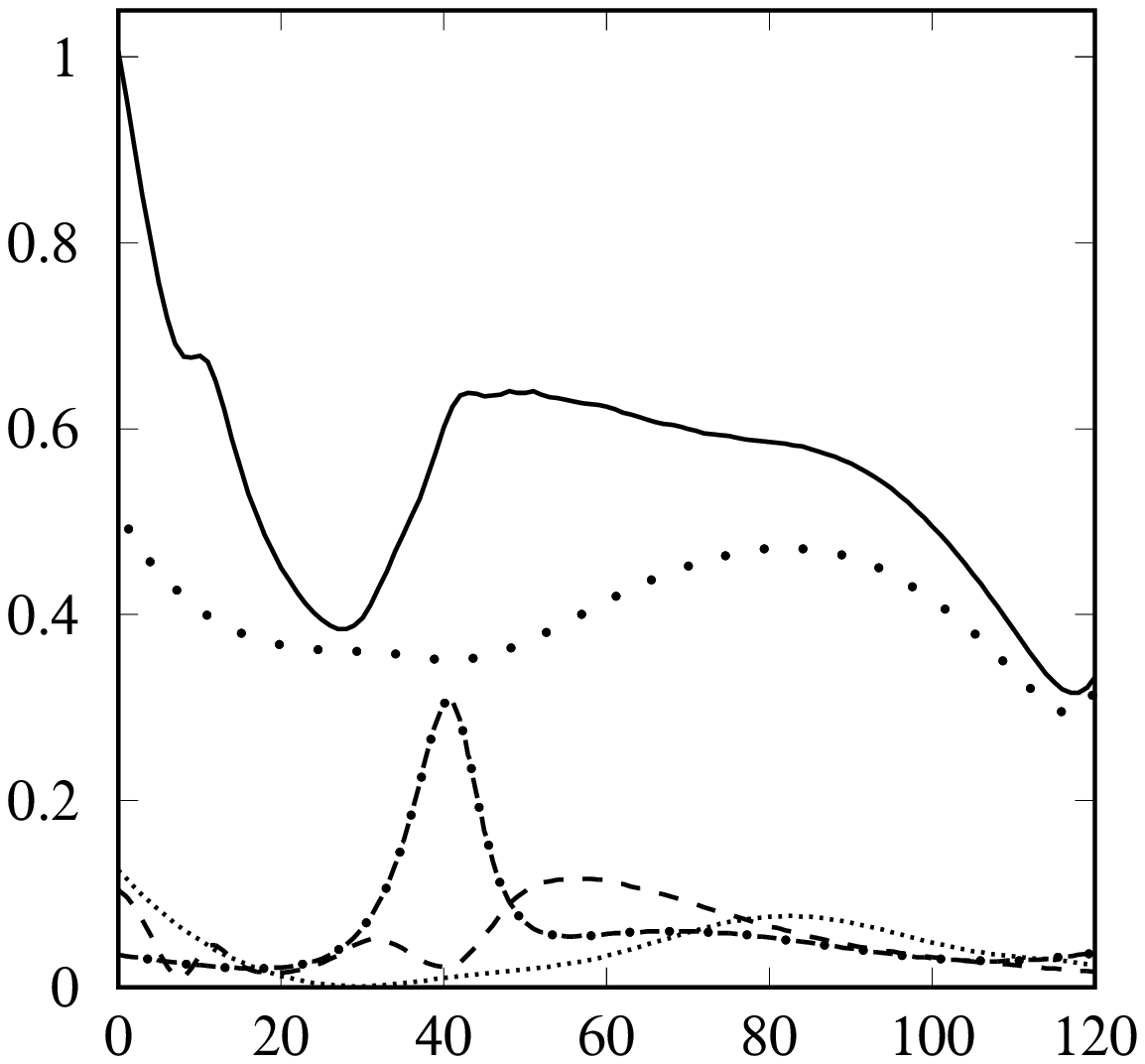}
\put(-13.0,13.0){\makebox(0,0) { \LARGE $  |T_{fi}|^{2}, 10^{-8} MeV^{-4} $ } }
\put(-5.0,2.0){\makebox(0,0){ \LARGE $ \theta_{2} $, \  deg. }}
}
%\caption
\begin{minipage}[h]{13cm}
{Figure 2: Dependence of squared moduli of the amplitudes  on the proton escape angle
$\theta_{2}$ for $p_{1} = 300 MeV/c, \theta_{1}=100^{\circ}, \phi_{1}=0^{\circ},
\phi_{2}=180^{\circ}, E_{\gamma}=360 MeV$ in the c.m. of the reaction.
The large dotted line is spectator model calculation, dash-dotted line is on-shell
$\pi N$ re-scattering, dashed line is off-shell $\pi N$ re-scattering, small dotted line is
$NN$ re-scattering, solid line includes all diagrams of Fig.1} 
\end{minipage}
%\label{f2}

\end{figure}
\suppressfloats[no-loc]

Double rescattering in the reaction (1) can be presented by diagrams in Fig.~2
and the similar diagrams obtained by permutating of identical nucleons in 
final state. 
We have calculated the contribution of these diagrams to the squared modulus
of the amplitude and found that in the kinematics under study 
($\Delta$--resonance region, large momenta of final nucleons) it amounts 
about 1\% of the contribution of the single rescattering.

Therefore we neglect double rescattering effects and the expression
for the amplitude of the reaction (1) is written as:
\begin{eqnarray}
 T\left(p_{1}, p_{2}, q, s,m_{s}; k,\lambda_{\gamma}, d,m_{d} \right)=
 T^{spect}\left(p_{1}, p_{2}, q, s,m_{s} ; k,\lambda_{\gamma}, d,m_{d} \right)+ & & \\
 +T^{\pi N}\left(p_{1}, p_{2}, q, s,m_{s}; k,\lambda_{\gamma}, d,m_{d} \right)+ 
 T^{NN}\left(p_{1}, p_{2}, q, s,m_{s} ; k,\lambda_{\gamma}, d,m_{d} \right) \ . \nonumber & & 
\end{eqnarray}
Usage of a $\pi$--meson on nucleon photoproduction amplitude evaluated in
\cite{david} provides a gauge invariance of the amplitude (18).
 
\section{}
\hspace*{\parindent} 
The amplitudes (18) are written in a mixed representation; polarization of
particles in initial state is described by a helicity of $\gamma$--quantum
$\lambda_{\gamma}$ and $z$-projection of a deuteron spin $m_{d}$, while 
a final state polarization is described by a total spin of a nucleon pair 
$s$ and its $z$-projection $m_{s}$.
To calculate polarization observables it is more convenient to use
helicity amplitudes \cite{nem}.
A transition to helicity amplitudes is carried out using a unitary 
transformation from a coupled basis of the nucleon pair to its helicity
basis and a transition from a projection of deuteron spin on $z$-axis,
(which coincides with a direction of $\gamma$-quantum momentum) to the
helicity of the deuteron:
\begin{eqnarray}
T\left(p_{1},\lambda_{1}, p_{2},\lambda_{2}, q ; k,\lambda_{\gamma}, d,\lambda_{d}\right) = 
\phantom{iiiiiiiiiiiiiiiiiiiiiiiiiiiiiiiiiiiiillllliiiii} & & \\
= \sum_{sm_{s}}\sum_{m_{1}m_{2}} \langle \frac{1}{2} m_{1} \frac{1}{2} m_{2} | s m_{s} \rangle
D^{*\frac{1}{2}}_{m_{1}\lambda_{1}}\left(\phi_{1},\theta_{1},0\right)
D^{*\frac{1}{2}}_{m_{2}\lambda_{2}}\left(\phi_{2},\theta_{2},0\right) 
(-)^{1-\lambda_{d}}T\left(p_{1}, p_{2}, q, s,m_{s} ; k,\lambda_{\gamma}, d,-\lambda_{d} \right) , & & \nonumber
\end{eqnarray}
where $D^{\frac{1}{2}}_{m_{i}\lambda_{i}}\left(\phi_{i},\theta_{i},0\right)$
are $D$--functions defined in \cite{nem}, 
$\phi_{i},\theta_{i}$ -- Euler angles, defining a transition to an
individual helicity system of a nucleon $i$.
The helicity amplitudes (19) are antisymmetric with respect to a permutation 
of final protons and they satisfy the relations following from the
parity conservation: 
\begin{eqnarray}
T\left(p_{1},\lambda_{1}, p_{2},\lambda_{2}, q ; k,\lambda_{\gamma}, d,\lambda_{d} \right)= 
\prod_{i}\eta_{i} \left(-\right)^{s_{i}-\lambda{i}}
T\left(p'_{1},-\!\lambda_{1}, p'_{2},-\!\lambda_{2}, q' ; k', -\!\lambda_{\gamma}, d',-\!\lambda_{d} \right) \ ,
& &
\end{eqnarray} 
where $\eta_{i},\ s_{i}$ are intrinsic parity and spin of a particle $i$,
the momenta in a right-hand side of the equation are obtained from the 
initial ones by a reflection in $xz$--plane. 
In a case of coplanar kinematics these relations coincide with those for
the two-particle reaction.
The reaction (1) has in total 24 helicity amplitudes and to define them
one has to know 47 independent observables (a common phase factor remains
undetermined). 
For the coplanar kinematics the relations (20) decrease by a factor two
the number of independent helicity amplitudes of the reaction (1) and their
number becomes the same as for the two-particle reaction 
$\gamma d \rightarrow p n$.

\begin{figure} [ht]
\unitlength=1cm
\hbox{
 \centering \includegraphics[width=1.0\textwidth]{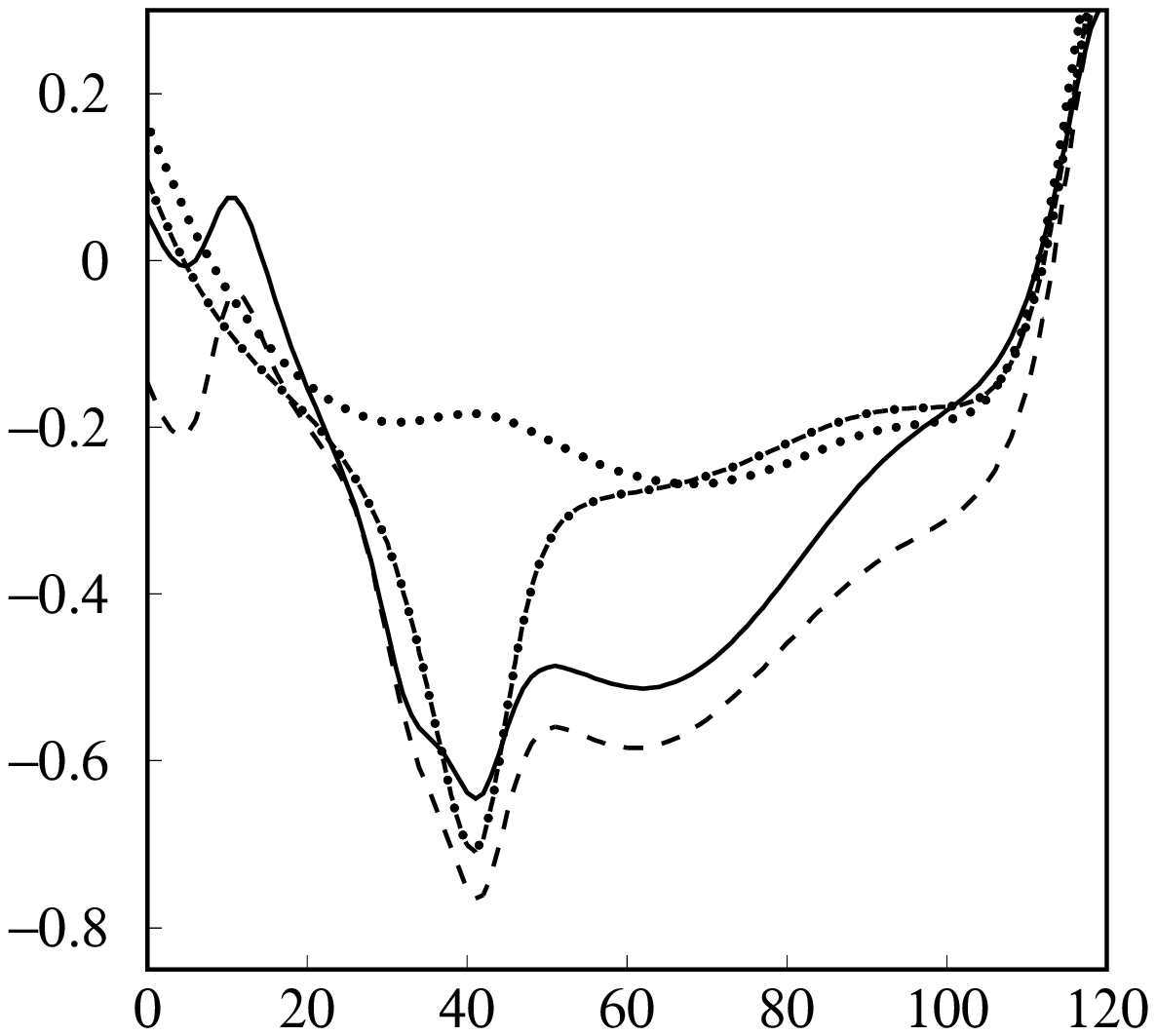}
\put(-15.0,12.0){\makebox(0,0) { \LARGE $  T_{00,20} $ } }
\put(-5.0,2.0){\makebox(0,0) {\LARGE $ \theta_{2} $, \  deg.} }
}
\begin{minipage}[h]{13.5cm}
 { Figure 3: Influence of the re-scattering effects on the tensor analyzing power 
  connected to the target polarization. Kinematics is the same as in Fig.2. The dotted
  line is spectator model calculation, dash-dotted line includes on-shell $\pi N$ re-scattering,
dashed line includes on-shell and off-shell $\pi N$ re-scattering, \  \  solid line includes 
on-shell and off-shell $\pi N$ re-scattering and $NN$ re-scattering.  } 
\end{minipage}
\end{figure}
\suppressfloats[no-loc]

A general expression for polarization observables of the three-particle 
reaction (1) is defined according to \cite{nem}:
\begin{eqnarray}
F^{I_{1}M_{1},I_{2}M_{2}}_{I_{\gamma}M_{\gamma},I_{d}M_{d}}=
\frac{SpT\tau_{I_{\gamma}M_{\gamma}}\tau_{I_{d}M_{d}}T^{\dagger}\tau_{I_{1}M_{1}}\tau_{I_{2}M_{2}}}{SpTT^{\dagger}} \ ,& &
\end{eqnarray}
where $T$ -- helicity amplitudes (19), $\tau_{I_{i}M_{i}}$ -- spherical
spin-tensor of a nucleon having a momentum $p_{i}$, 
$\tau_{I_{\gamma}M_{\gamma}}, \tau_{I_{d}M_{d}}$ -- spherical spin-tensors
of $\gamma$--quantum and deuteron respectively. 
One has to note that for a given energy of the initial state polarization
observables of a two-particle reaction are functions of the single kinematic
variable -- scattering angle, while for a three-particle reaction they
are functions of five variables and their choice is ambiguous.
Often for such variables one takes momentum and two escape angles of one
particle and two escape angles of other particle.

\section{}
\hspace*{\parindent}

To demonstrate an influence of re-scattering effects we present the results of
calculation of the squared moduli of amplitudes corresponding to the diagrams
in Fig. 1, analyzing powers of the reaction (1) connected to beam 
polarization $T_{22,00}$, to target polarization  $T_{00,20}$ and 
to polarization of one of the final protons $P1_{y}$.
The calculations were performed in a center-of-mass frame of the reaction (1) 
for the coplanar kinematics.
These parameters are shown as functions of proton escape angle $\theta_{2}$
for fixed values of the following five variables:
$p_{1}=300 \ MeV, \ \theta_{1}=100^{\circ}, \ \phi_{1}=0^{\circ}, \ \phi_{2}=180^{\circ}$ 
and $\gamma$ -- quantum energy $E_{\gamma } = 360 \ MeV $.

Fig. 3 it is shown a dependence of squared moduli of the amplitudes 
averaged over the spins of initial particles and summed over the spins of
final particles.
For the proton escape angle $\theta_{2}$ larger than $120^{\circ}$ the 
contribution of the re-scattering is small if compared to that of the
spectator mechanism, therefore the dependence is shown only in the 
$\theta_{2}$ range between $0^\circ$ and $120^{\circ}$.
The squared modulus of the ``on-shell'' $\pi N$ -- re-scattering amplitude 
shows a most characteristic behavior. 
It has a clear maximum in the region $ \theta_{2} \sim 40^{\circ}$.
This maximum comes from the first term of the amplitude (11) 
corresponding to the case when a proton with a momentum $p_{1}$ takes 
part in the $\pi N$ -- re-scattering.
In this kinematic region the value of $|p_{-}|$ in the first term
of (11) is small and the integral (15) reaches the maximum.
The main contribution to the ``off-shell'' $\pi N$ -- re-scattering comes
from the first term of the amplitude (12) which, like in the previous case,
corresponds to the re-scattering of a proton with the momentum $p_{1}$.
The behavior of the ``off shell" $\pi N$ -- re-scattering amplitude is defined
mainly by a dependence of the integral
\begin{eqnarray}
I_{off} = P \int d^{3}p' \frac{u_{0}\left(\frac{1}{2}\left(|d-2p'|\right)\right)}{q'^{2}-m^{2}_{\pi}} &  & \nonumber
\end{eqnarray}
on the proton escape angle $\theta_{2}$.
The contribution of the nucleon-nucleon re-scattering increases drastically
with decreasing a relative kinetic energy of final nucleons because  
a scattering phase in ${}^{1}S_{0}$ state grows.
In Fig.3 this corresponds to the increase of the contribution of the 
nucleon-nucleon re-scattering in a region of small $\theta_{2}$. 
A contribution of the ${}^{1}S_{0}$--state re-scattering decreases
with increasing the relative kinetic energy, but in the same time 
a contribution of the scattering in the states with orbital momenta
$ L=1,2 $ : $ {}^{3}P_{0}$,$ {}^{3}P_{1}$,$ {}^{3}P_{2}$,$ {}^{1}D_{2}$
grows.

\begin{figure} [ht]
\unitlength=1cm
\hbox{
 \centering \includegraphics[width=1.0\textwidth]{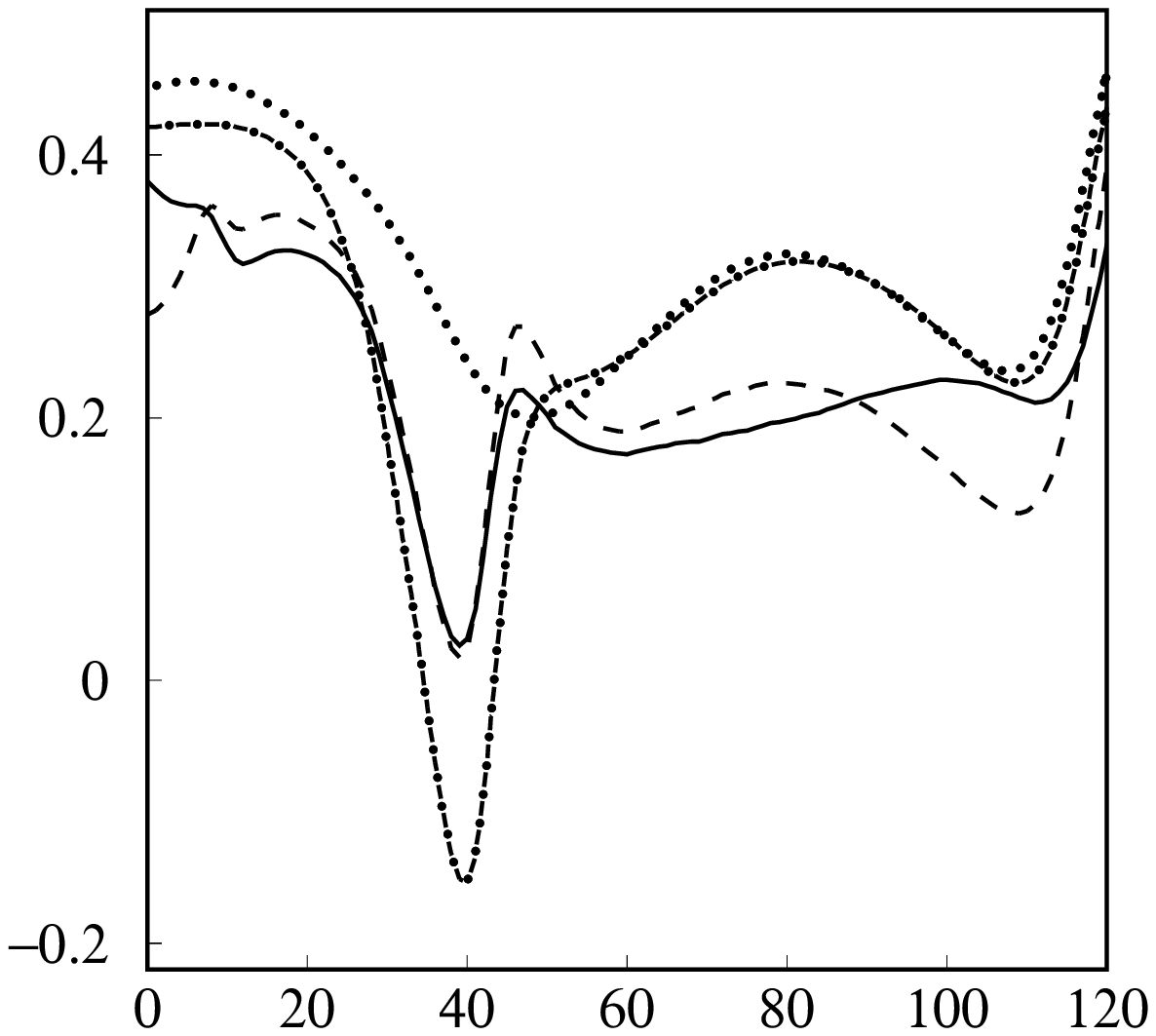}
\put(-15.0,12.0){\makebox(0,0) { \LARGE $  T_{22,00} $}}
\put(-5.0,2.0){\makebox(0,0) {\LARGE $ \theta_{2} $, \  deg.} }
}
\begin{minipage}[h]{13.5cm}
 { Figure 4:
\ Influence of the re-scattering effects on the tensor analyzing power connected to the 
beam polarization. Kinematics is the same as in Fig.2, the curve defenitions are the same as in
Fig.3. } 
\end{minipage}
\end{figure}
\suppressfloats[no-loc]

In Fig.2 an increase of a contribution of the nucleon-nucleon re-scattering
in a region $\theta_{2} \sim  90^{\circ}$ is determined by a re-scattering in 
$P-$ and $D-$ states. 

In Figs. 3,4 is shown an influence of the re-scattering on the tensor 
analyzing power connected to the target polarization, 
$T_{00,20}=F^{00,00}_{00,20}$, and the one connected to the beam 
polarization, $T_{22,00}=F^{00,00}_{22,00}$.
The major contribution comes from the ``on shell"  $\pi N$ -- re-scattering
which leads to a deep minimum in the region $ \theta_{2} \sim 40^{\circ} $. 
When the ``off shell" $\pi N$ -- re-scattering and $NN$ -- re-scattering are
taken into account the depth of the minimum decreases, this is especially 
true for $T_{22,00}$.
Note that a decrease of the depth of the minimum in $ T_{22,00}$ is 
connected with the `off shell"  $\pi N$ -- re-scattering, although it is
in the region $\theta_{2} \sim 40^{\circ}$ that squared modulus of its
amplitude has a minimum.
The major contribution to the decrease of the depth of the minimum of
$ T_{22,00}$ comes from an interference between ``of-shell'' and ``on-shell''
$\pi N$--re-scattering amplitudes.
 
In Fig.5 one can see an influence of the re-scattering on the polarization
of a final proton having a momentum $p_{1}$ for an unpolarized initial state,
which is defined as:
\begin{eqnarray} 
 P1_{y}=\frac{SpTT^{\dagger}\sigma_{y}(1)}{SpTT^{\dagger}} \ , & & \nonumber
\end{eqnarray}
where $T$ are helicity amplitudes (19), 
\  ${\displaystyle \frac{1}{2}\sigma_{y}(1)}$ -- operator of a 
$y$--projection of the proton spin, other components of the polarization 
vector in coplanar kinematics are equal to zero.
One can see that the re-scattering effects noticeably changes the behavior
of $ P1_{y}$, especially in a region $\theta_{2} \sim 40^{\circ} - 60^{\circ}$.
Note that in the reaction (1) with unpolarized initial state the proton 
polarization is non-zero even in the spectator model without accounting for
$\pi N$ ¨ $NN$ -- re-scattering.

\begin{figure} [ht]
\unitlength=1cm
\hbox{
 \centering \includegraphics[width=1.0\textwidth]{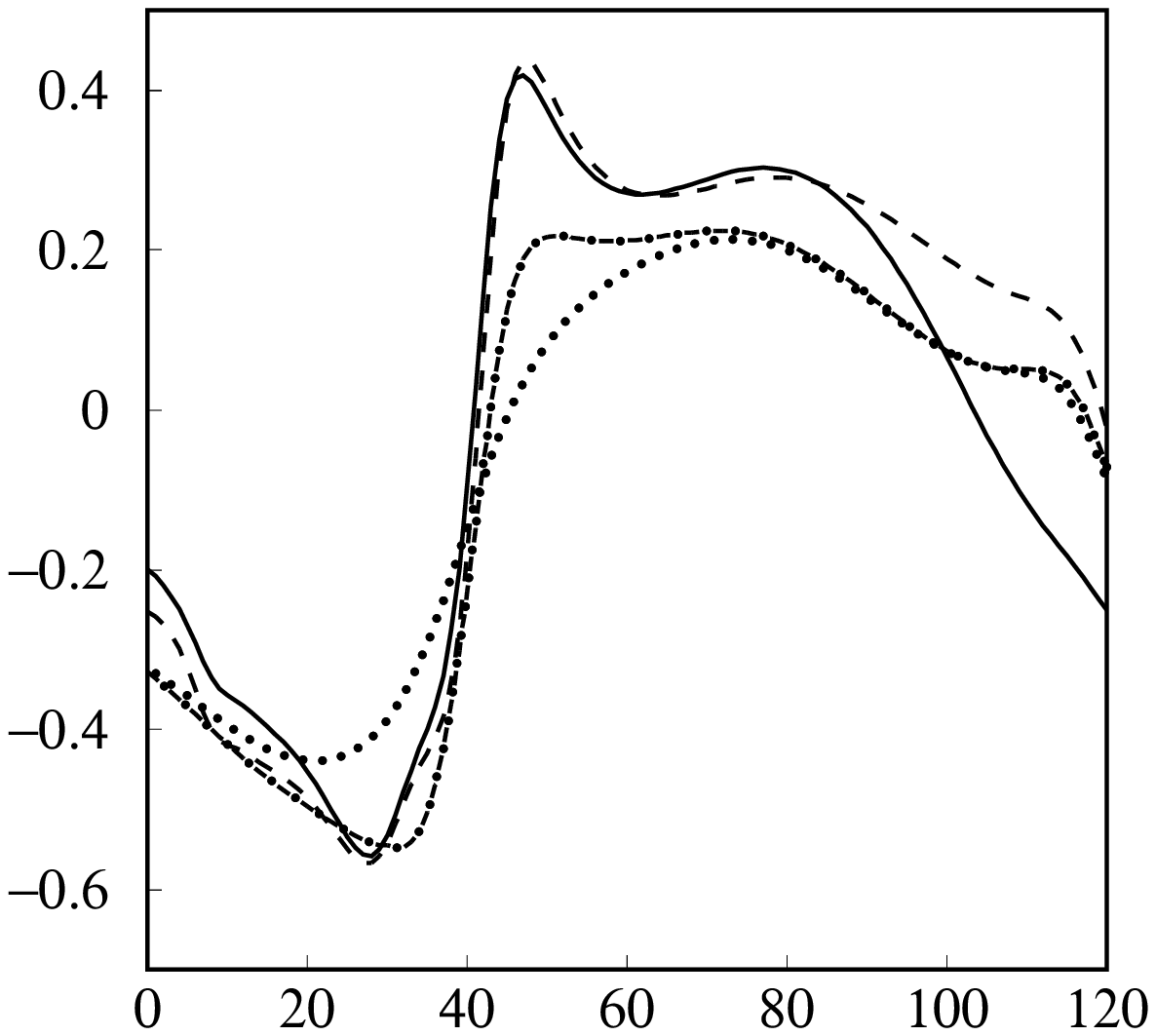}
\put(-15.0,12.0){\makebox(0,0) { \LARGE $ P1_{y}$ } }
\put(-5.0,2.0){\makebox(0,0) {\LARGE $ \theta_{2} $, \  deg.} }
}
\begin{minipage}[h]{13.5cm}
 { Figure 5:
 Influence of the re-scattering effects on the nucleon polarization.  Kinematics and 
 the curve definitions are the same as in Fig.4. } 

\end{minipage}
\end{figure}
\suppressfloats[no-loc]

This can be explained by a necessity to include besides Born terms a 
contribution of $\Delta$ -- isobar in $s$-- and $u$-- channel as well as
$\rho$--meson and $\omega$--meson exchange in $t$--channel for the description
of the photoproduction of $\pi$--meson on nucleon.
An account of the contribution of $\Delta$ -- isobar in $s$ -- channel leads
to the appearance of an imaginary part in the amplitude of the photoproduction
of $\pi$--meson on nucleon, and as a consequence, to the polarization of a
final nucleon with an unpolarized initial state.
In the same time for the reaction $e^{-}d \rightarrow e'^{-}pn$ the polarization
of final nucleons with an unpolarized initial state turns out to be non-zero
only if one takes into consideration $pn$ -- re-scattering \cite{recl},
because the $eN$--re-scattering amplitude is real in Born approximation.

Calculation of the influence of pion-nucleon and nucleon-nucleon re-scattering
on polarization observables of the reaction $\gamma d \rightarrow pp \pi^{-}$,
performed in the framework of diagrammatic approach, has shown that the
effects of re-scattering play a noticeable role in the behavior of the
polarization observables in the kinematic region of $\Delta$-isobar with
large momenta of protons in the final state.
Contribution of the re-scattering effects must be taken into account in 
the analysis of experimental data.

\end {document}